\documentclass[12pt]{article}
\usepackage{fullpage,graphicx,psfrag,amsmath,amsfonts,verbatim}
\usepackage[small,bf]{caption}
\usepackage{amssymb}
\graphicspath{{figures/}}

\usepackage{listings}
\usepackage{longtable}
\usepackage{subcaption}
\usepackage{booktabs}

\usepackage{color}

\definecolor{codegreen}{rgb}{0,0.6,0}
\definecolor{codegray}{rgb}{0.5,0.5,0.5}
\definecolor{codepurple}{rgb}{0.58,0,0.82}
\definecolor{backcolour}{rgb}{0.95,0.95,0.98}

\lstdefinestyle{mystyle}{ backgroundcolor=\color{backcolour},
      keywordstyle=\color{magenta}, numberstyle=\tiny\color{codegray},
      stringstyle=\color{codepurple}, basicstyle=\ttfamily\small,
      breakatwhitespace=false, breaklines=true, captionpos=t, keepspaces=true,
      numbers=left, numbersep=5pt, showspaces=false, showstringspaces=false,
      showtabs=false, tabsize=2, }

\definecolor{seagreen}{rgb}{0.18, 0.55, 0.34}
\definecolor{mediumviolet-red}{rgb}{0.78, 0.08, 0.52}
\definecolor{khaki}{rgb}{0.94, 0.9, 0.55}

\newcommand{\BEAS}{\begin{eqnarray*}}
    \newcommand{\EEAS}{\end{eqnarray*}}
    \newcommand{\BEA}{\begin{eqnarray}}
    \newcommand{\EEA}{\end{eqnarray}}
    \newcommand{\BEQ}{\begin{equation}}
    \newcommand{\EEQ}{\end{equation}}
    \newcommand{\BIT}{\begin{itemize}}
    \newcommand{\EIT}{\end{itemize}}
    \newcommand{\BNUM}{\begin{enumerate}}
    \newcommand{\ENUM}{\end{enumerate}}

    \newcommand{\BA}{\begin{array}}
    \newcommand{\EA}{\end{array}}

    \newcommand{\eg}{{\it e.g.}}
    \newcommand{\ie}{{\it i.e.}}

    \newcommand{\ones}{\mathbf 1}

    \newcommand{\reals}{{\mbox{\bf R}}}

    \newcommand{\Expect}{\mathop{\bf E{}}}

    {\begin{quote}}{\end{quote}}

    \makeatletter
    \long\def\@makecaption#1#2{
       \vskip 9pt
       \begin{small}
       \setbox\@tempboxa\hbox{{\bf #1:} #2}
       \ifdim \wd\@tempboxa > 5.5in
            \begin{center}
            \begin{minipage}[t]{5.5in}
            \addtolength{\baselineskip}{-0.95pt}
            {\bf #1:} #2 \par
            \addtolength{\baselineskip}{0.95pt}
            \end{minipage}
            \end{center}
       \else
        \hbox to\hsize{\hfil\box\@tempboxa\hfil}
       \fi
       \end{small}\par
    }
    \makeatother

    \newcounter{oursection}

    \newcounter{lecture}

\usepackage{mathtools}
\usepackage{adjustbox}
\usepackage{booktabs}
\usepackage{csquotes}

\usepackage{hyperref}
\hypersetup{ colorlinks   = true,    
urlcolor     = blue,    
linkcolor    = blue,    
citecolor    = red      
}

\title{Simple and Effective Portfolio Construction\\ with Crypto Assets}
\author{Kasper Johansson \and Stephen Boyd }
\begin{document}
\maketitle

\begin{abstract}
We consider the problem of constructing a portfolio that combines traditional
financial assets with crypto assets.  We show that despite the documented
attributes of crypto assets, such as high volatility, heavy tails, excess
kurtosis, and skewness, a simple extension of traditional risk allocation
provides robust solutions for integrating these emerging assets into broader
investment strategies. Examination of the risk allocation holdings suggests an
even simpler method, analogous to the traditional 60/40 stocks/bonds allocation,
involving a fixed allocation to crypto and traditional assets, dynamically
diluted with cash to achieve a target risk level.
\end{abstract}

\clearpage
\tableofcontents
\clearpage

\section{Introduction}
Since the introduction of cryptocurrencies in 2009~\cite{nakamoto2008peer}, the
field of crypto trading has rapidly grown. In this note we consider the problem
of constructing a portfolio of assets, including a combination of traditional
financial assets such as stocks and bonds with crypto assets such as Bitcoin and
Ethereum.   Our conclusion is that despite the well known extreme swings of
crypto currency values, simple standard portfolio construction methods suffice
to realize the benefits of including crypto assets in a portfolio.

\begin{figure}
\centering
\includegraphics[width=0.7\textwidth]{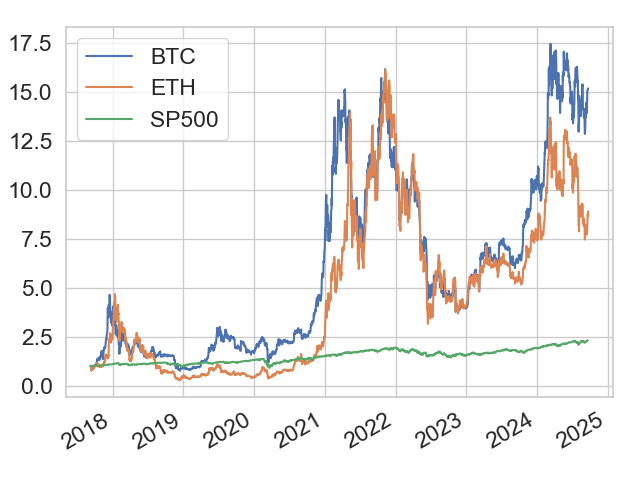}
\caption{Normalized prices of BTC, ETH, and SP500.}
\label{fig:asset_values}
\end{figure}

Figure~\ref{fig:asset_values} shows the normalized prices of Bitcoin (BTC),
Ethereum (ETH), and the S\&P~500 index (SP500) over the past six years. To the
eye it reasonably seems that crypto returns are quite different from traditional
returns.  
Table~\ref{tab:performance_metrics} lists some metrics for these asset returns
over the same period, which also suggest that crypto asset returns fundamentally
differ from traditional asset returns. For example, ETH at one point dropped in
value by a factor of almost $20\times$, while the maximum drop in value of SP500
is only one third.
\begin{table}
\centering
\caption{Performance metrics for BTC, ETH, and SP500.}
\begin{tabular}{lrrr}
\toprule
Metric & BTC & ETH & SP500 \\
\midrule
Return (\%) & 43.3 & 47.0 & 13.7 \\
Volatility (\%) & 58.1 & 71.6 & 19.5 \\
Sharpe & 0.75 & 0.66 & 0.70 \\
Drawdown (\%) & 83.3 & 93.9 & 33.9 \\
\bottomrule
\end{tabular}
\label{tab:performance_metrics}
\end{table}

\begin{figure}
\centering
\includegraphics[width=0.5\textwidth]{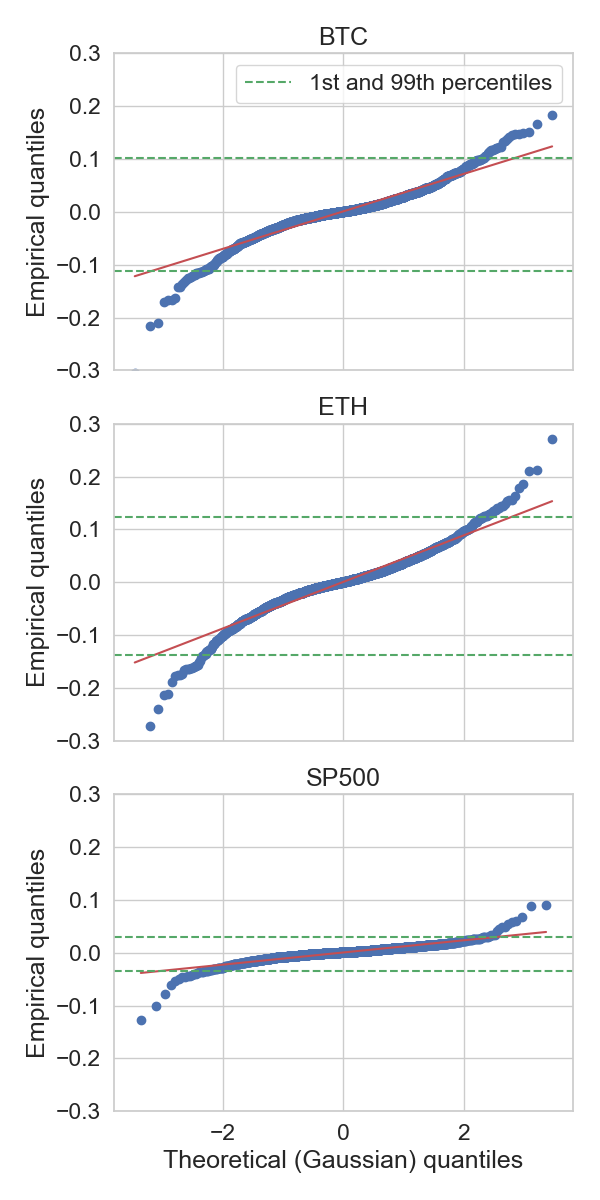}
\caption{Quantile-quantile plot of log returns of BTC, ETH, and SP500.}
\label{fig:QQ_plot}
\end{figure}

Stylized facts of crypto returns, such as extreme volatility, heavy tails,
excess kurtosis, and skewness, have been well documented~\cite{liu2021risks,
hu2019cryptocurrencies, karim2024exploring}. Figure~\ref{fig:QQ_plot} shows
quantile-quantile (QQ) plots of the log returns of BTC, ETH, and
SP500 over the last six years.
All three return distributions have tails that deviate significantly from the
normal distribution, with the crypto asset returns exhibiting more extreme tail
behavior than the market index. The 1st and 99th return percentiles are shown in
the plots as dashed green lines; here too we see that crypto asset returns have
considerably bigger tails than the market index, even when normalized to have
the same volatility.

The documented characteristics of crypto asset returns have made conventional
investors hesitant to include them in their portfolios, due to perceptions of
risk and unpredictability. While asset managers may have additional concerns,
such as legislative risk and other fundamental factors like the ongoing debate
about the legitimacy of crypto as a real asset~\cite{yarow2013krugman,
smith2013bitcoin, krugman2011golden, ciolli2013bitcoin}, we will focus here on
the concerns related to the statistical properties of crypto asset returns.

These characteristics of crypto asset returns have also led to a wealth of
research on how to extend traditional portfolio construction methods to include
crypto assets, including methods similar to those used for traditional
assets~\cite{holovatiuk2020cryptocurrencies, brauneis2019cryptocurrency,
burggraf2019risk, hu2019modelling}, as well as more complex machine learning
driven methods~\cite{ruiz2022cryptocurrencies, jiang2017cryptocurrency,
lucarelli2020deep, jiang2017deep, ramkumar2021cryptocurrency,
ruiz2022cryptocurrencies}. 

Despite the documented differences between crypto and conventional asset
returns, some authors have argued that the two asset classes are fundamentally
similar~\cite[Chap.~2]{palomar2024portfolio}, even though the crypto asset
returns and volatilities are much higher. We agree with this perspective. In this
note we show a simple method for constructing a portfolio of traditional and
crypto assets using a risk allocation framework, (hopefully) debunking the idea
that novel and complex machine learning approaches are necessary to manage a
portfolio that includes crypto assets.
Based on a back-test of the risk allocation method, we propose an even
simpler portfolio construction method, reminiscent of the traditional
60/40 stocks/bonds split, which consists of a 90/10 split of traditional and crypto 
assets, followed by dynamic (time-varying) dilution with cash, to achieve a 
given ex-ante risk.  We refer to this simple portfolio as DD90/10.

\paragraph{Outline.}
In \S\ref{s-related} we review previous and related work. 
We introduce in \S\ref{s-cra}
an approach for constructing a portfolio that combines
traditional and crypto assets within a risk allocation framework. We illustrate
the performance of this method on historical data in \S\ref{s-results}. Finally,
in~\S\ref{s-DD90/10} we propose the DD90/10 portfolio allocation strategy, and show
that it has performance similar to the risk allocation strategy.

\section{Related work}\label{s-related}

\subsection{Stylized facts of financial return data}
Stylized facts of financial return series refer to a set of empirical
observations that are consistently observed across different financial markets
and asset classes. We review some of the most prominent stylized facts for
financial assets, and refer the reader to~\cite[Chap.~2]{palomar2024portfolio}
for a more comprehensive overview.

\paragraph{Equities.}
Stylized facts for equity returns include the non-normal distribution of
returns, characterized by heavy tails and excess
kurtosis~\cite{malmsten2010stylized, bulla2006stylized}. Another fact is
volatility clustering, where large price movements tend to be followed by
additional large movements, regardless of direction. While returns themselves
are generally uncorrelated over time, the absolute or squared returns often
display strong autocorrelation, highlighting a pattern in the magnitude of
fluctuations. Furthermore, the leverage effect is a notable feature, where
negative returns increase future volatility to a greater extent than positive
returns of the same size. Equity return distributions have also been shown to
exhibit asymmetry between positive and negative returns~\cite{jiang2020stock}.

\paragraph{Cryptocurrencies.}
Cryptocurrencies exhibit several patterns similar to traditional financial
assets, but with more extreme behavior~\cite{hu2019cryptocurrencies,
ghosh2023return}. They are highly volatile, with large price fluctuations and
heavy-tailed return distributions. Volatility tends to cluster, with periods of
high volatility followed by more volatility. While returns show no
autocorrelation in the short term, return magnitudes do exhibit autocorrelation.
Cryptocurrencies also exhibit asymmetry in returns~\cite{karim2024exploring}.

\subsection{Portfolio construction}
Portfolio construction involves selecting a combination of assets by balancing
the trade-off between expected return and portfolio risk. Here we discuss key
historical contributions to the field, along with some recent advancements. For
more extensive overviews, refer to texts such as
\cite[Chap.~14]{grinold2000active}, \cite[Chap.~6]{narang2024inside}, and the
studies by \cite{cornuejols2006optimization, kolm201460,
chakrabarty2023mathematical, gunjan2023brief, boyd2024markowitz}. 

\paragraph{Markowitz portfolio construction.}
Before Markowitz's seminal work in 1952, portfolio construction was largely
based on heuristics and rules of thumb. Markowitz introduced a quantitative
framework for portfolio construction, where the return of a portfolio is modeled
as a random variable, and the expected return is maximized for a desired level
of risk~\cite{markowitz1952portfolio}. Despite its simplicity, and having over
70 years of history, the Markowitz model remains the foundation of quantitative
investing to this day~\cite{boyd2024markowitz}.

Extensions of the Markowitz model include the Black-Litterman
model~\cite{black1990asset, black1992global, black1989universal}, fully flexible
views~\cite{meucci2010fully}, and conditional value-at-risk (CVaR)
optimization~\cite{rockafellar2000optimization}, to name a few. The
Black-Litterman model is a Bayesian approach to portfolio construction, where
the prior distribution is based on the equilibrium market portfolio, and the
posterior distribution is updated with user-specific views on the expected
returns of the assets. Fully flexible views is a generalization of
Black-Litterman, allowing for nonlinear views on the returns. CVaR optimization
replaces the variance in the Markowitz model with the conditional value-at-risk
of the portfolio, penalizing the tail risk of the portfolio, and directly
addresses the issue of a non-normal return distribution.

\paragraph{Machine learning based portfolio construction.}
Typically, portfolio construction is split into two parts: data modelling and
portfolio optimization~\cite[Chap.~1]{palomar2024portfolio}. Data modelling is
concerned with predicting the expected returns and covariances of the assets,
and portfolio optimization concerns selecting a portfolio of assets that trades
off expected return and risk. However, with the growing popularity of machine
learning, and documented criticism of the Markowitz
model~\cite{michaud1989markowitz_enigma}, in recent years several studies have
proposed machine learning based portfolio construction methods as alternatives
to this traditional framework~\cite{gurrib2022machine, bartram2021machine}. In
particular, it has become popular to combine the two parts of portfolio
construction into a single end-to-end machine learning model, where market
features are fed into the model, and the model outputs a trade list. The
argument for this has been that splitting portfolio construction into two parts
is suboptimal, as there are uncertainties in the return forecasts that are not
accounted for in the portfolio optimization component of portfolio
construction~\cite[\S 5]{kelly2023financial}. Although theoretically appealing,
we have yet to see a wide-spread adoption of these methods in practice, and the
Markowitz model remains the dominant framework for portfolio
construction~\cite{boyd2024markowitz}. We refer the reader to~\cite[\S
5]{kelly2023financial} for a detailed review of end-to-end machine learning
models for portfolio construction.

\paragraph{Risk-based portfolio construction.} Risk-based portfolio construction
methods rely on estimates of the asset return covariances, but do not require an
estimate of the expected returns of the assets, making them attractive for
practitioners who do not have access to data sources for estimating expected
returns reliably. A trivial example of a risk-based portfolio construction
method is the equally weighted portfolio. Another popular portfolio is the
minimum variance portfolio, which is also the mean-variance efficient portfolio
when expected returns are equal. Other risk-based portfolio construction methods
include risk parity~\cite{Qian119}, where the risk contribution of each asset is
equal, and maximum diversification portfolios~\cite{choueifaty2008toward}. These
portfolios can all be computed via convex optimization~\cite[\S
4.4]{johansson2023covariance}, which makes them reliable, fast, and
practical~\cite{boydconvex}. These portfolio construction methods can be
implemented in just a few lines of domain specific languages for convex
optimization such as CVXPY~\cite{cvxpy}.

\paragraph{Risk and covariance estimation.}
Risk-based portfolio construction methods rely on estimates of the portfolio
risk. There are in general two ways to estimate portfolio risk. The first is to
use a realized measure of the variance of the portfolio. There are many such
methods, including the exponentially weighted moving average (EWMA), methods
based on mean absolute deviation or the rolling median~\cite{geary1936moments,
geary1947testing}, as well as autoregressive conditional heteroskedasticity
(ARCH) and generalized ARCH (GARCH) models~\cite{engle1982autoregressive,
bollerslev_1986, engle1986modelling}. The second way to estimate portfolio risk
is to leverage a covariance matrix of the asset returns. The literature on
covariance estimating in finance is vast, but probably the most popular method
is to use an iterated covariance matrix~\cite{DCC, cov_barrat_2022}. This method
decomposes the covariance $\Sigma$ as $\Sigma = V R V$, where $V$ is
a diagonal matrix with the asset standard deviations on the diagonal and $R$ is
the correlation matrix. Typically, $V$ and $R$ are estimated separately, using
EWMAs with different half-lives. We will refer to this method as the iterated
EWMA (IEWMA)~\cite[\S 2.5]{johansson2023covariance}. It is also possible to
dynamically adjust the half-lives of the EWMAs, to account for time-varying
market conditions~\cite{johansson2023covariance}.

\subsection{Crypto trading}
Several methods to managing portfolios of cryptocurrencies have been proposed,
and these tend to be separated from investment strategies proposed for
traditional assets.

\paragraph{Diversification.} Many studies have noted that the returns of
cryptocurrencies are uncorrelated with traditional
assets~\cite{bakry2021bitcoin, yermack2024bitcoin,
holovatiuk2020cryptocurrencies}. This means that they can be leveraged to
diversify a portfolio of traditional assets, and thus increase the risk-adjusted
return of the portfolio~\cite{holovatiuk2020cryptocurrencies, bakry2021bitcoin}.

\paragraph{Portfolio construction.} 
Much of the literature on cryptocurrency trading is focused on machine learning
or deep learning~\cite{ruiz2022cryptocurrencies}.
In~\cite{jiang2017cryptocurrency} an end-to-end convolutional neural network,
taking in raw price data and outputting a trade list, is proposed. A deep
Q-learning portfolio management framework is proposed
in~\cite{lucarelli2020deep}. The authors of~\cite{jiang2017deep} introduce a
financial-model-free reinforcement learning framework, incorporating
convolutional neural networks, recurrent neural networks, and long short-term
memory models. In~\cite{ramkumar2021cryptocurrency} ARIMA models, convolutional
neural networks, and long short-term memory methods are used for cryptocurrency
price forecasting and multiple portfolio construction strategies are evaluated
with the forecasted prices as input. For a more detailed review of machine
learning in cryptocurrency portfolio management, see \eg,~\cite[\S
4.2]{ruiz2022cryptocurrencies}.

Some crypto studies use traditional portfolio construction methods, such as the
Markowitz framework. In~\cite{holovatiuk2020cryptocurrencies}, the authors show
that crypto assets can improve the performance of a mean-variance optimized
portfolio. The authors of~\cite{brauneis2019cryptocurrency} show how to
implement a mean-variance optimized portfolio of cryptocurrencies and that it
outperforms an equally weighted portfolio as well as single cryptocurrencies.
The paper~\cite{giudici2020network} proposes an extension of the Markowitz
model, combining random matrix theory and network measures to manage a portfolio
of crypto assets. In~\cite{platanakis2019portfolio} the authors design crypto
portfolios using variance-based constraints in the Black-Litterman model, to
account for estimation uncertainties. 

Other studies focus on risk-based portfolio construction methods, \ie, those
that do not rely on an estimate of the expected returns of the assets.
In~\cite{burggraf2019risk} multiple risk-based portfolio construction methods
are evaluated, and the authors find that most of them outperform single
cryptocurrencies and the equally weighted portfolio. The authors
of~\cite{hu2019modelling} find that minimizing the variance and conditional
value-at-risk of a portfolio of cryptocurrencies, yields a portfolio that
outperforms the market.

\section{Constrained risk allocation}\label{s-cra} The most common approach to
portfolio construction is to formulate the problem as a trade-off between
expected return and risk, as suggested by Markowitz in
1952~\cite{markowitz1952portfolio}. The main practical challenge with this
framework is that it requires estimating expected returns of the assets. These
are very difficult to estimate and, for obvious reasons, successful estimation
techniques are proprietary; large hedge funds and asset managers have entire
teams dedicated to estimating expected returns, using data sources for which
they pay large premiums~\cite{palomar2024portfolio}. Here we describe a risk
allocation approach, which does not require an estimate of the expected returns
of the assets. It is based on the idea of risk parity~\cite{Qian119}, with
additional constraints on the portfolio weights and a risk limit. We refer to
this method as constrained risk allocation (CRA).

\subsection{Constrained risk allocation problem}
We consider a portfolio of $n$ non-cash assets, plus cash. We denote the asset
weights as $w\in \reals^n$, with $w \geq 0$, where $w_i$ is the fraction of the
total portfolio value held in asset $i$. We let $c\geq 0$ denote the fraction of
the portfolio value held in cash, so we have $\ones^Tw + c =1$, where $\ones$ is
the vector with all entries one. We refer to $\ones^Tw$ as our asset exposure.

Let $\Sigma$ denote the $n \times n$ (estimate of the) covariance matrix of the
returns.  Assuming cash is risk-free, the portfolio risk is $w^T\Sigma w$.  (The
volatility is the squareroot of this.) Inspired by the identity
\[
w^T\Sigma w = \sum_{i=1}^n w_i (\Sigma w)_i,
\]
we define the risk contribution of asset $i$ as $w_i (\Sigma w)_i$. In risk
allocation we specify the fraction of risk to be held in each asset, as the
vector $\rho \in \reals^n$, with $\rho > 0$ and $\ones^T \rho=1$. We interpret
$\rho_i$ as the fraction of total portfolio risk contributed by asset $i$.  
(We assume that all entries of $\rho$ are positive; if any were zero, we simply
would not include that asset in the portfolio.) Thus we have
\BEQ\label{e-risk-alloc} w_i(\Sigma w)_i = \rho_i w^T\Sigma w, \quad i=1,
\ldots, n. \EEQ The special case $\rho = (1/n)\ones$ corresponds to risk parity,
where all assets contribute an equal fraction of the risk.

The CRA problem is defined as \BEQ \label{eq:CRA}
\begin{array}{ll}
\mbox{minimize}   &  c \\
\mbox{subject to} & w \geq 0, \quad c\geq 0, \quad \ones^Tw + c =1\\ 
& w_i (\Sigma w)_i = \rho_i w^T \Sigma w, \quad i=1, \ldots, n\\
& w^T \Sigma w \leq \sigma^2, \quad Fw \leq g,
\end{array}
\EEQ where $w \in \reals^n$ and $c\in \reals$ are the variables, $\sigma^2$ is
the maximum allowed risk, and $F\in \reals^{m \times n}$ and $g\in \reals^m$
describe constraints on the portfolio. In words: We choose the portfolio to
minimize cash holdings (or equivalently, maximize asset exposure), subject to a
given risk allocation, a total risk limit, and some additional constraints on
the weights.

We will assume that $g>0$ and $F\geq 0$ (elementwise), with each row of $F$
nonzero. We will soon see that this implies the CRA problem \eqref{eq:CRA}
always has a unique solution. The weight constraints can be used to enforce a
maximum weight on each asset, or a maximum weight on a subset of assets, \eg,
crypto assets.

\subsection{Solution via convex optimization} \label{s-convex-solution} The CRA
problem \eqref{eq:CRA} is not itself a convex optimization problem, but it can
be solved efficiently via convex optimization. We first consider the risk
allocation constraints \eqref{e-risk-alloc} alone, together with $w \geq 0$. It
can be shown that $w$ satisfies these constraints if and only if it has the form
\[
w = \alpha x^\star,
\]
where $\alpha \geq 0$ and $x^\star\in\reals^n$ is the unique solution of the
convex optimization problem \BEQ\label{eq:risk_allocation_opt} \mbox{minimize}
\quad (1/2)x^T\Sigma x - \sum_{i=1}^n \rho_i \log x_i, \EEQ with variable $x \in
\reals^n$ (and implicit constraint $x>0$). See, \eg,~\cite[\S
17]{cvx_book_additional}. Thus the set of weights that satisfy the risk
allocation constraints is a ray, with a direction that can be found by solving a
convex optimization problem.

Now we take this very specific form for $w$ and substitute it back into the
original CRA problem \eqref{eq:CRA}, dropping the risk allocation constraints
and $w \geq 0$ since they are automatically satisfied. This gives us the problem
\[
\begin{array}{ll}
\mbox{minimize}   &  c \\
\mbox{subject to} & \alpha \geq 0, \quad c\geq 0, \quad \alpha \ones^Tx^\star + c =1\\ 
& \alpha^2 (x^\star)^T \Sigma x^\star \leq \sigma^2, \quad \alpha Fx^\star \leq g,
\end{array}
\]
with scalar variables $\alpha$ and $c$. Note that the quantities
$\ones^Tx^\star$, $(x^\star)^T\Sigma x^\star$, and $Fx^\star$ are constants in
this problem.

We can solve this simple problem analytically. Minimizing $c$ is the same as
maximizing $\alpha$.  Along with $\alpha \geq 0$, all constraints on $\alpha$
are (positive) upper limits:
\[
\alpha \leq \frac{1}{\ones^T x^\star}, \quad
\alpha \leq \frac{\sigma}{\left((x^\star)^T\Sigma x^\star\right)^{1/2}}, \quad
\alpha \leq \frac{g_i}{(Fx^\star)_i}, \quad i= 1, \ldots, m.
\]
(Each of the denominators is positive due to our assumptions and $x^\star>0$.)
It follows that the solution is \BEQ\label{eq:CRA_solution} \alpha^\star =
\min\left\{ \frac{1}{\ones^T x^\star},~ \frac{\sigma}{\left((x^\star)^T\Sigma
x^\star\right)^{1/2}},~ \frac{g_i}{(Fx^\star)_i}, \quad i= 1, \ldots, m
\right\}. \EEQ Roughly speaking: Scale the unconstrained risk allocation weights
as large as possible with all constraints holding.

\paragraph{Summary.} The two step solution procedure is summarized as follows: 
\begin{enumerate}
\item Solve the optimization problem~\eqref{eq:risk_allocation_opt} to obtain
$x^\star$.
\item The unique solution of the CRA problem is then given by $w^\star =
\alpha^\star x^\star$ where $\alpha^\star$ is given by~\eqref{eq:CRA_solution}.
\end{enumerate}

We note that Feng and Palomar have suggested a more sophisticated formulation of
the CRA problem, in which the risk allocations need only hold
approximately~\cite{feng2015scrip, wu2020general}, 
\cite[Chap.~11]{palomar2024portfolio}. 
This formulation can be approximately solved by
solving a sequence of convex problems. Our simple formulation is, however, good
enough for us to make our main point.

\paragraph{Variation.} We can modify the way $\alpha^\star$ is computed
in~\eqref{eq:CRA_solution}. Instead of estimating the standard deviation of the
unconstrained risk allocation portfolio $x^\star$ as $\left((x^\star)^T \Sigma
x^\star\right)^{1/2}$, we compute the realized return trajectory of the
portfolio $x^\star$, \ie, $(x^\star)^T r_t$, $t=1,2, \ldots$, where $r_t$ is the
vector of asset returns at time $t$. We then compute an estimate of the standard
deviation of the portfolio return trajectory, using, \eg, a EWMA.  Thus to
compute the scaling (which sets the cash dilution) we directly estimate the
standard deviation of the return trajectory with the unconstrained risk
allocation weights, rather than find it from our estimated covariance matrix
(which we use to compute the risk allocation weights $x^\star$ in step~1.) We
have found that leads to a modest but significant improvement in portfolio
performance.

\section{CRA results}\label{s-results}
\subsection{Data and experimental setup}
\begin{figure}
\centering
\includegraphics[width=0.7\textwidth]{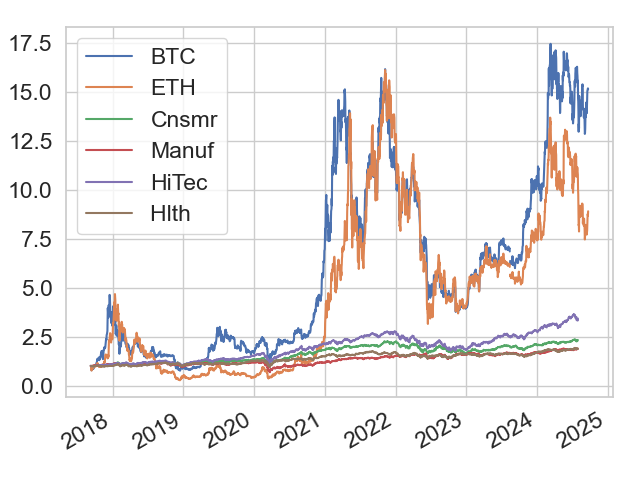}
\caption{Normalized prices of BTC, ETH, and four industry portfolios.}
\label{fig:all_prices}
\end{figure}

\paragraph{Data.} We consider daily close prices of two crypto assets, and ETH,
with data from LSEG Data and Analytics.  We also consider four daily traded
industry portfolios: consumer goods and services, manufacturing and utilities,
technology and communications, as well as healthcare, medical equipment, and
drugs; these were obtained from Kenneth French's data
library~\cite{french_data_lib}. The data spans from September 8th, 2017, to September
22nd, 2024, for a total of 2565 days, or 1729 trading days. (Although crypto assets are traded
every day, we rebalance our portfolios only on market trading days; we do,
however, realize gains and losses on crypto assets on weekends and holidays.)
Figure~\ref{fig:all_prices} shows the normalized price evolution of the six
assets, and table~\ref{tab:asset_metrics} lists some metrics for them. The data
and code to reproduce the results are available at
\begin{center}
\url{https://github.com/cvxgrp/crypto_portfolio}.
\end{center}

\begin{table}
\centering
\caption{Performance metrics for the six assets.}
\begin{tabular}{lcccccc}
\toprule
Metric & BTC & ETH & Cnsmr & Manuf & HiTec & Hlth \\
\midrule
Return (\%) & 43.5 & 47.1 & 14.1 & 11.4 & 20.7 & 10.8 \\
Volatility (\%) & 58.1 & 71.6 & 19.3 & 20.5 & 23.9 & 18.0 \\
Sharpe & 0.73 & 0.60 & 0.76 & 0.58 & 0.90 & 0.62 \\
Drawdown (\%) & 83.3 & 93.9 & 28.5 & 42.7 & 35.4 & 26.8 \\
\bottomrule
\end{tabular}
\label{tab:asset_metrics}
\end{table}

\begin{figure}
\centering
\includegraphics[width=0.7\textwidth]{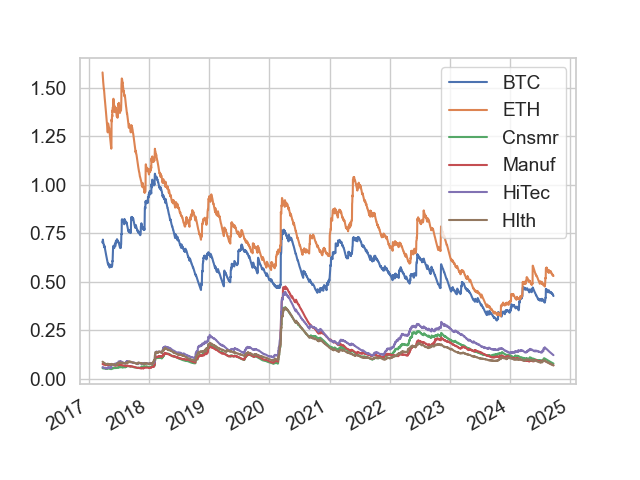}
\caption{Estimated annualized volatilities of the six assets.}
\label{fig:volatility}
\end{figure}

\begin{figure}
\centering
\begin{subfigure}{0.49\textwidth}
\centering
\includegraphics[width=\textwidth]{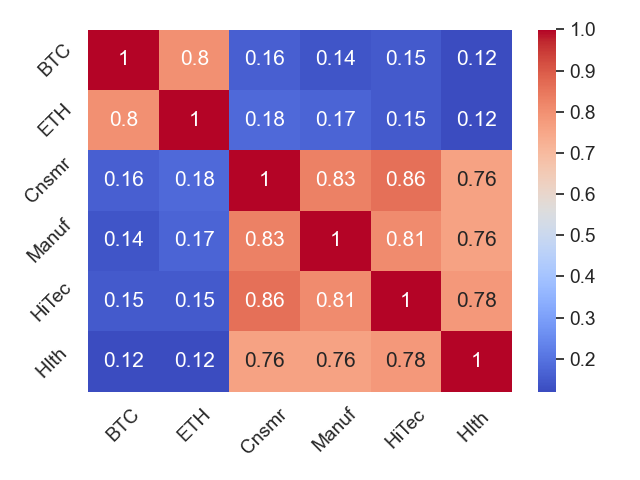}
\caption{March 1st, 2020.}
\label{fig:correlation_2020}
\end{subfigure}
\hfill
\begin{subfigure}{0.49\textwidth}
\centering
\includegraphics[width=\textwidth]{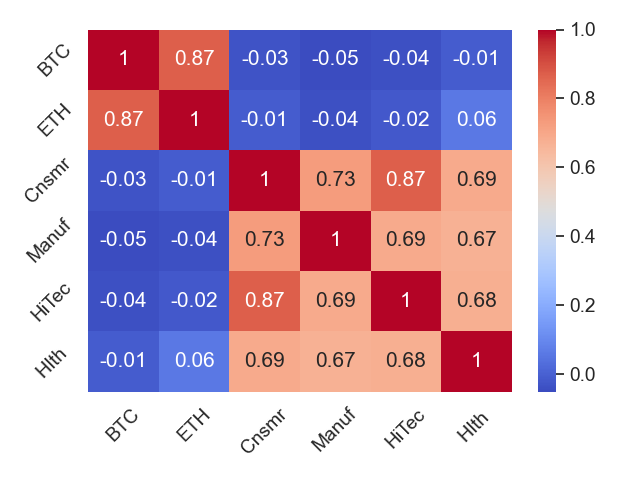}
\caption{May 1st, 2023.}
\label{fig:correlation_2023}
\end{subfigure}
\caption{Estimated correlation matrices on two different dates.}
\label{fig:covariance_matrix}
\end{figure}

\paragraph{Risk model.} We estimate the covariance matrix of the assets using an
iterated EWMA, described in detail in~\cite[\S2.5]{johansson2023covariance}. We use a 63-day half-life for the volatility
estimate and a 125-day half-life for the correlation estimate. The estimated
volatilities of the six assets are shown in figure~\ref{fig:volatility}.
Two examples (from different time-periods) of the estimated correlation matrix
are shown in figure~\ref{fig:covariance_matrix}.

\paragraph{Simulation and parameters.} We simulate three portfolios. \BIT
\item \emph{Industries} contains only the four industry portfolios and cash.
\item \emph{Crypto} contains only the two crypto assets and cash.
\item \emph{Combined} contains all six assets and cash. \EIT We rebalance the
portfolios every trading day, using risk parity. We impose a 10\% annualized
risk limit on the portfolio, \ie, $\sigma = 0.1 \sqrt D$, where $D=250$ is the
number of trading days in a year. To estimate the risk of the unconstrained risk
parity portfolio $x^\star$, we use a 10-day half-life EWMA of the portfolio
return. We also impose a 10\% maximum weight constraint on crypto assets, \ie,
for BTC and ETH combined. These limits were chosen as reasonable values that one
might use in practice; the results are not sensitive to these choices.

\subsection{Metrics}
We describe the metrics used to evaluate the performance of the portfolios over
the time interval $t=1, \ldots, T$.

\paragraph{Return.} The (realized) return of the portfolio at time $t$ is given
by
\[
w_t^T r_t,
\]
where $r_t$ and $w_t$ are the vector of (realized) asset returns and the
portfolio weights at time $t$, respectively. The annualized (realized) return is
given by
\[
\bar r = \frac{D}{T} \sum_{t=1}^T w_t^T r_t.
\]

\paragraph{Volatility.} The (realized) annualized volatility of the portfolio is
given by
\[
\left(\frac{D}{T} \sum_{t=1}^{T}\left(r_t-\overline r\right)^2\right)^{1/2}.
\]

\paragraph{Sharpe ratio.} 
The Sharpe ratio is the ratio of the annualized return to the annualized
volatility. 

\paragraph{Drawdown.} Let $V_t$ denote the portfolio value in time period $t$,
starting from $V_1=1$, with returns compounded or re-invested. These are found
from the recursion $V_{t+1} = (1+w_t^Tr_t) V_t$, $t=1,\ldots, T-1$. The
(maximum) drawdown of the portfolio is defined as 
\[
\max_{1 \leq t_1 < t_2\leq T} \left(1- \frac{V_{t_2}}{V_{t_1}} \right),
\]
\ie, the maximum fractional drop in value form a previous high.

\subsection{Results}
\begin{figure}
\centering
\begin{subfigure}{0.7\textwidth}
\centering
\includegraphics[width=\textwidth]{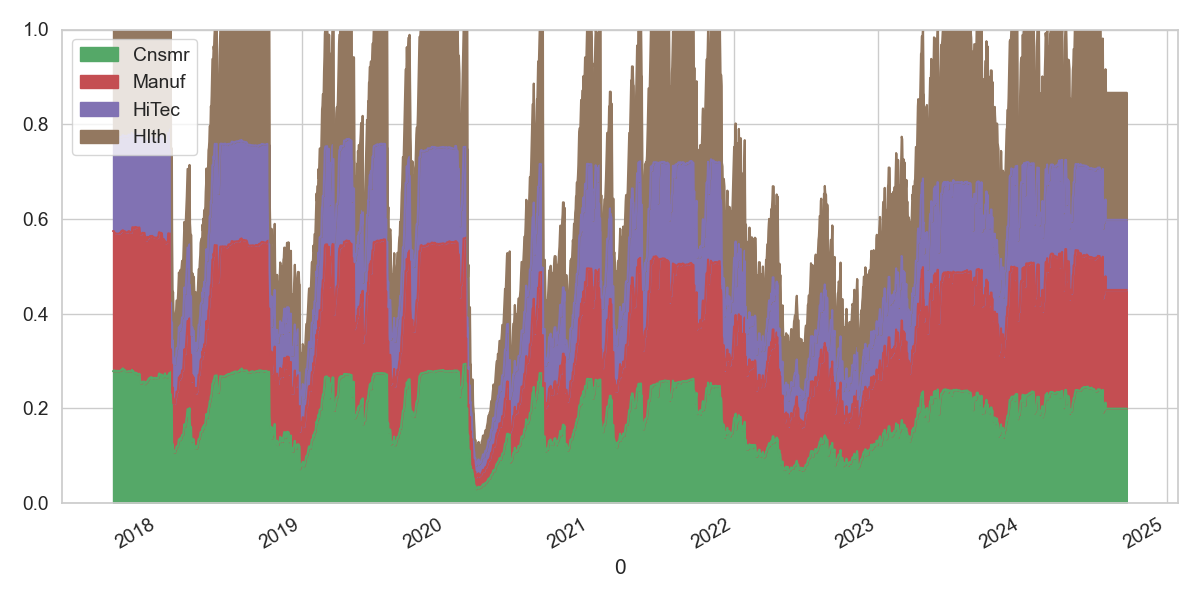}
\caption{Industry portfolio.}
\label{fig:portfolio_weights_industry}
\end{subfigure}
\hfill
\begin{subfigure}{0.7\textwidth}
\centering
\includegraphics[width=\textwidth]{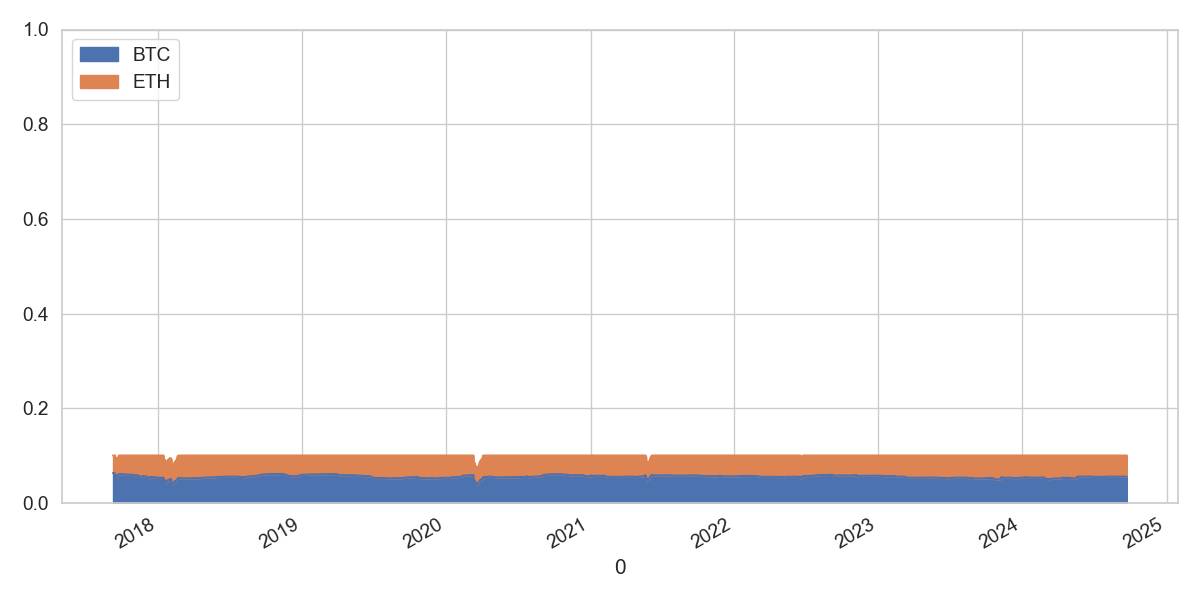}
\caption{Crypto portfolio.}
\label{fig:portfolio_weights_crypto}
\end{subfigure}
\begin{subfigure}{0.7\textwidth}
\centering
\includegraphics[width=\textwidth]{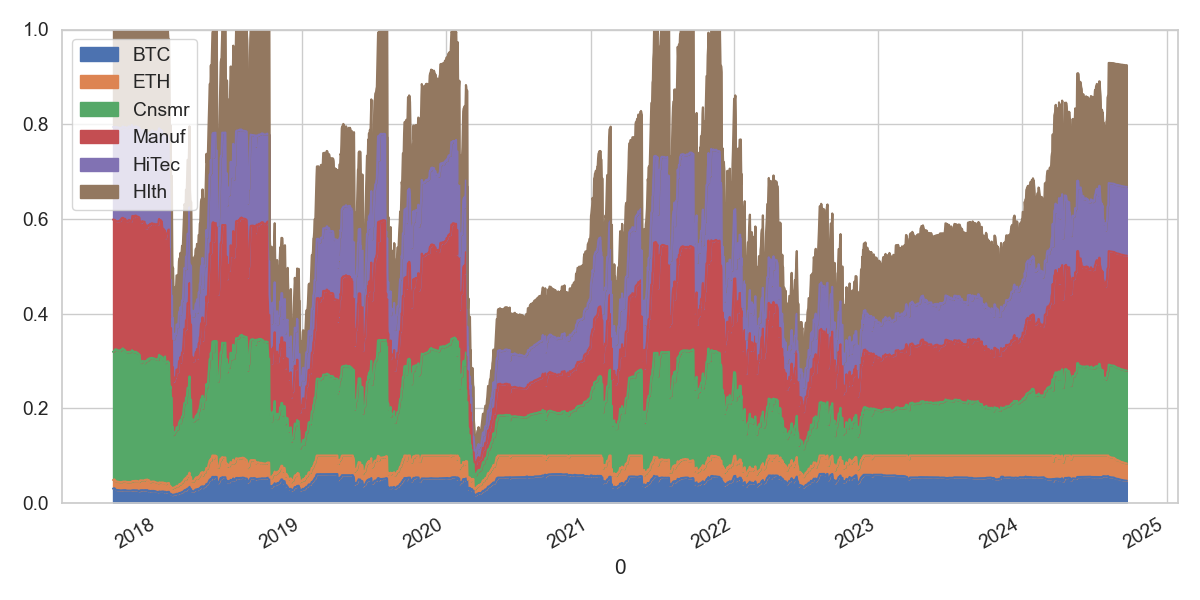}   
\caption{Combined portfolio.}
\label{fig:portfolio_weights_all}
\end{subfigure}
\caption{Portfolio weights of the three portfolios.  The cash weight is shown as
uncolored.}
\label{fig:portfolio_weights}
\end{figure}

\paragraph{Portfolio weights.} The portfolio weights of the three portfolios are
shown in figure~\ref{fig:portfolio_weights}.
As expected, the crypto portfolio holds mostly cash, due to the 10\% crypto
limit. The industry and combined portfolios have more diversified weights,
varying over time. As expected  these portolios hold a lot of cash during the
turbulent 2020 period. On average, the industry portfolio holds 25\% cash, the
crypto portfolio 90\% cash, and the combined portfolio 33\% cash.

\begin{figure}
\centering
\includegraphics[width=0.7\textwidth]{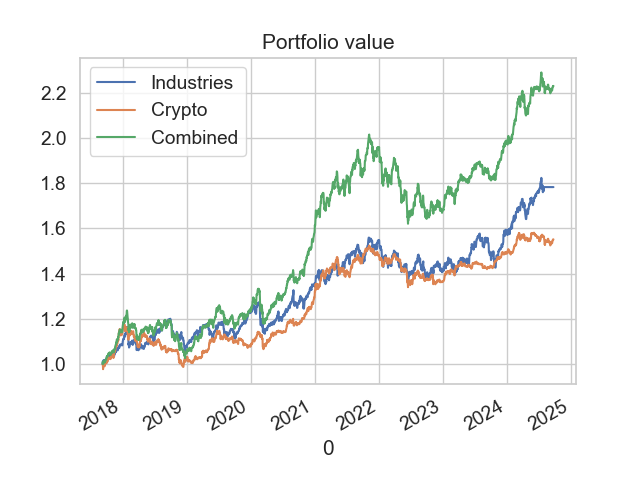}
\caption{Portfolio values of the three portfolios.}
\label{fig:portfolio_values}
\end{figure}

\begin{figure}
\centering
\includegraphics[width=0.75\textwidth]{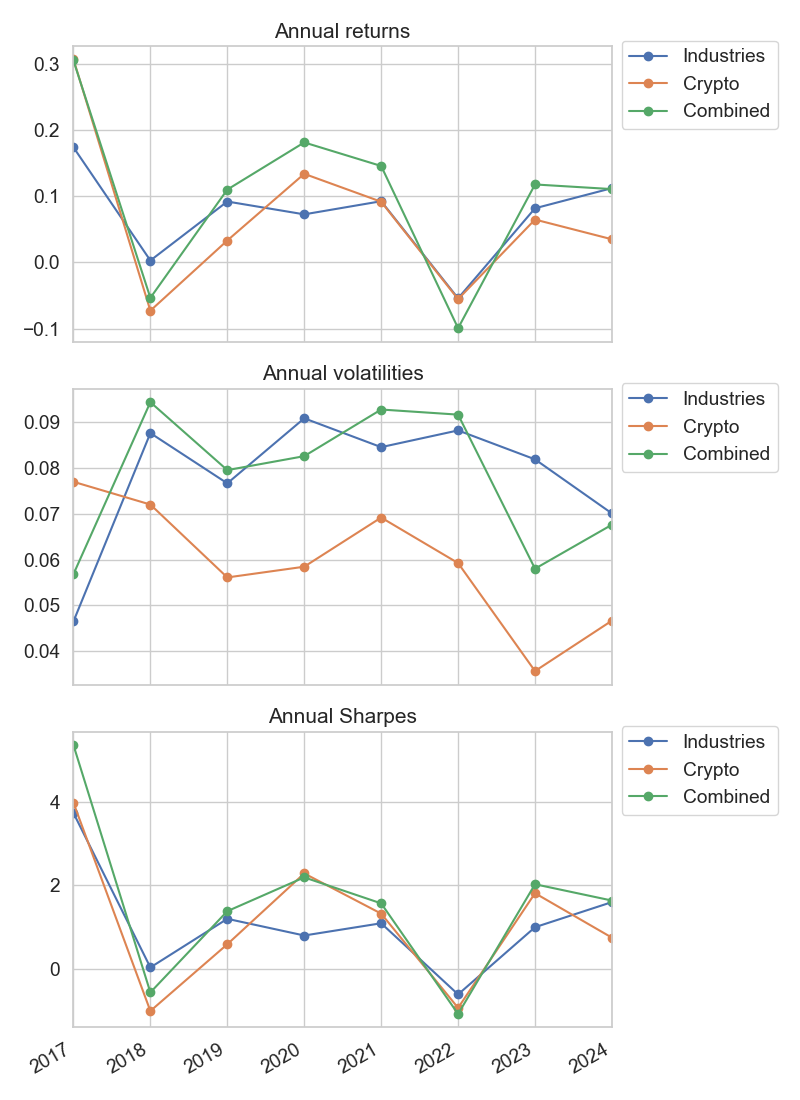}
\caption{Annual performance metrics of the three portfolios.}
\label{fig:annual_metrics}
\end{figure}

\paragraph{Performance.} Figure~\ref{fig:portfolio_values} shows the value of
the three portfolios over time, normalized to one at the start of the
simulation. (SR denotes Sharpe ratio.)
The aggregate performance of the three portfolios is shown in
table~\ref{tab:portfolio_stats}.
\begin{table}
\centering
\caption{Portfolio performance metrics.}
\begin{tabular}{lrrr}
\toprule
Metric             & Industries & Crypto & Combined    \\ 
\midrule
Return (\%)   & 6.0       & 4.5         & 8.2    \\
Volatility (\%)    & 8.2       & 6.0         & 8.2    \\
Sharpe       & 0.73     & 0.75       & 1.00  \\
Drawdown (\%)  & 12.5      & 15.9        & 19.6   \\ 
\toprule
\end{tabular}
\label{tab:portfolio_stats}
\end{table}
Including crypto assets in the portfolio clearly gives a noticeable boost,
despite allocating less than 10\% of the portfolio to them. The combined
portfolio has a higher return and Sharpe ratio than the industry portfolio,
while having similar volatility and drawdown. To get a better understanding of
the performance differences, figure~\ref{fig:annual_metrics} shows the return,
volatility, and Sharpe ratio of the three portfolios over each year of the
simulation.

\clearpage
\subsection{Shapley attributions}
We would like to attribute the performance of the portfolio to the different
asset classes. Shapley values account for each assets's contribution to the
portfolio, ensuring a fair allocation. They are uniquely characterized by
satisfying a collection of desirable properties, including fairness,
monotonicity, and full attribution~\cite{shapley1953value,
huettner2012axiomatic, zhang2023did, fryer2021shapley, owen2017shapley, moehle2021portfolio}. We will
now look at the Shapley attribtutions to each industry and to crypto as
a whole, for the combined portfolio. The Shapley attributions of the different
asset classes for the combined portfolio are shown in table~\ref{tab:shapley}.
\begin{table}
\centering
\caption{Shapley attributions by asset category.}
\begin{tabular}{lrrrrr|r}
\toprule
& {Cnsmr} & {Manuf} & {HiTec} & {Hlth} & {Crypto} & {Total} \\
\midrule
{Return (\%)} & 1.8 & 0.7 & 2.5 & 0.8 & 2.5 & 8.2 \\
{Volatility (\%)} & 1.8 & 1.8 & 2.0 & 1.9 & 0.7 & 8.2 \\
{Sharpe} & 0.20 & 0.06 & 0.26 & 0.08 & 0.40 & 1.0 \\
{Drawdown (\%)} & 3.3 & 2.4 & 3.2 & 3.5 & 7.3 & 19.6 \\
\bottomrule
\end{tabular}
\label{tab:shapley}
\end{table}
Crypto assets have the highest attribution to return and Sharpe. All assets
other than crypto have around a 2\% contribution to volatility; crypto has a
noticeably lower volatility contribution. Crypto assets have the highest
contribution to drawdown.

\section{Dynamically diluted 90/10 portfolio}\label{s-DD90/10}

\begin{figure}
\centering
\includegraphics[width=0.7\textwidth]{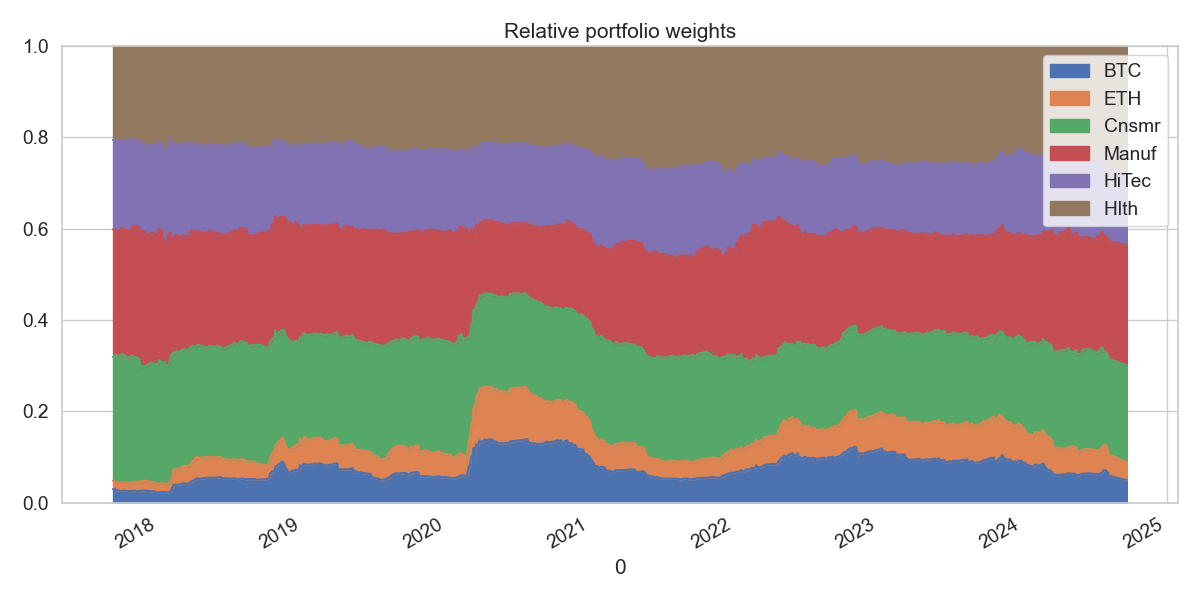}
\caption{Relative weights of the combined portfolio.}
\label{fig:relative_weights}
\end{figure}

Figure~\ref{fig:relative_weights} shows the relative non-cash weights, \ie,
$w/\ones^Tw$ for the combined portfolio over time.
We see that, apart from 2020, the relative weights are relatively stable and
evenly distributed with about 10\% in crypto assets (equally split between BTC
and ETH), and 90\% roughly equally split between the four industry portfolios.
This motivates an even simpler portfolio construction method, akin to the
popular 60/40 stocks/bonds allocation. \BIT
\item \emph{90/10 portfolio.} Construct a portfolio consisting of 90\% equities
(\eg, the four industries with equal weights) and 10\% crypto (\eg, equally
split between BTC and ETH).
\item \emph{Dynamic cash dilution.} Based on an estimate of the recent
volatility of the 90/10 portfolio, dilute the 90/10 portfolio with cash to
achieve the target risk $\sigma$ and respect weight limits. We refer to this
portfolio as the dynamically diluted 90/10 (DD90/10) portfolio. \EIT

\subsection{DD90/10 results}
\begin{figure}
\centering
\includegraphics[width=0.9\textwidth]{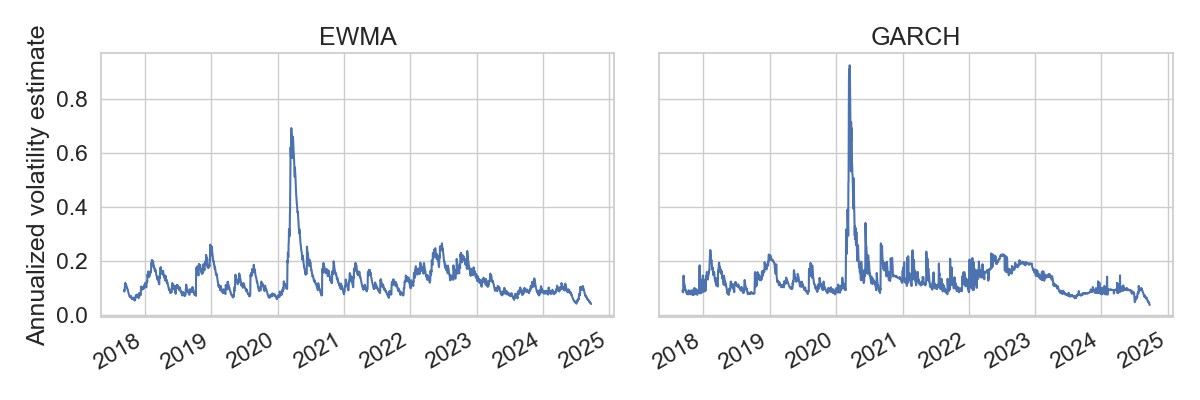}
\caption{Annualized volatility estimates of the 90/10 portfolio.}
\label{fig:volatility_estimates}
\end{figure}

\paragraph{Volatility estimators.} We evaluate the performance of the DD90/10
portfolio using two different volatility estimators: a 10-day half-life EWMA,
and a GARCH(1,1) model refitted every day on the last 250 days of data, using
the arch package in Python~\cite{sheppard2024arch}. (We tried several other
volatility estimators; all gave similar results.) The volatility
estimates of the 90/10 portoflio
are shown in figure~\ref{fig:volatility_estimates}.

\begin{figure}
\centering
\begin{subfigure}{0.7\textwidth}
\centering
\includegraphics[width=\textwidth]{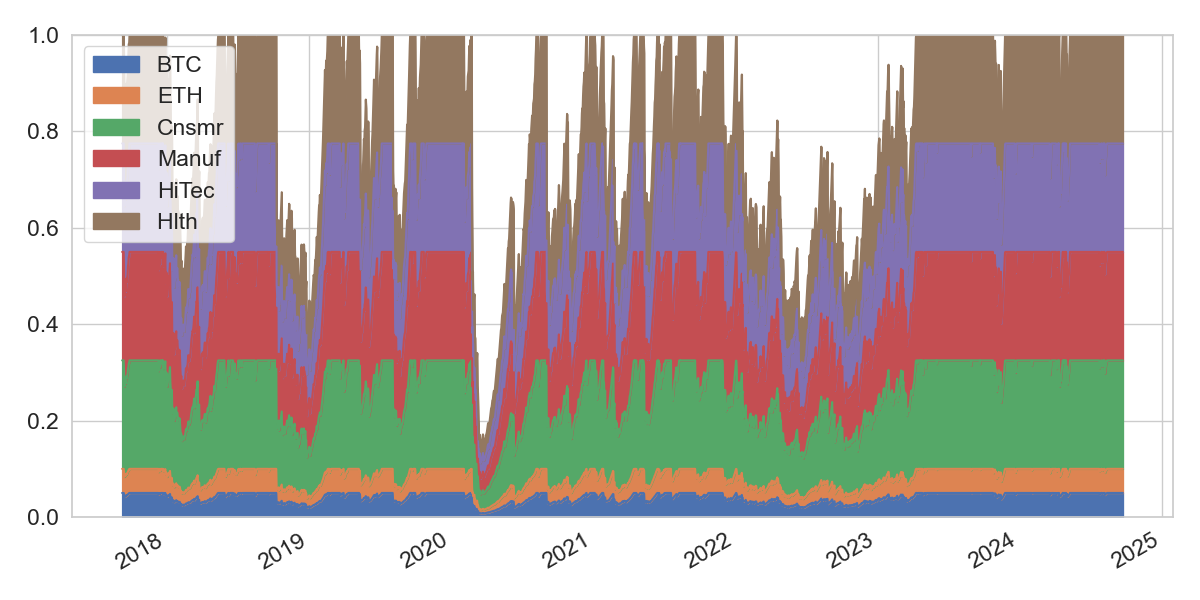}
\caption{EWMA volatility estimator.}
\label{fig:weights_ewma}
\end{subfigure}
\hfill
\begin{subfigure}{0.7\textwidth}
\centering
\includegraphics[width=\textwidth]{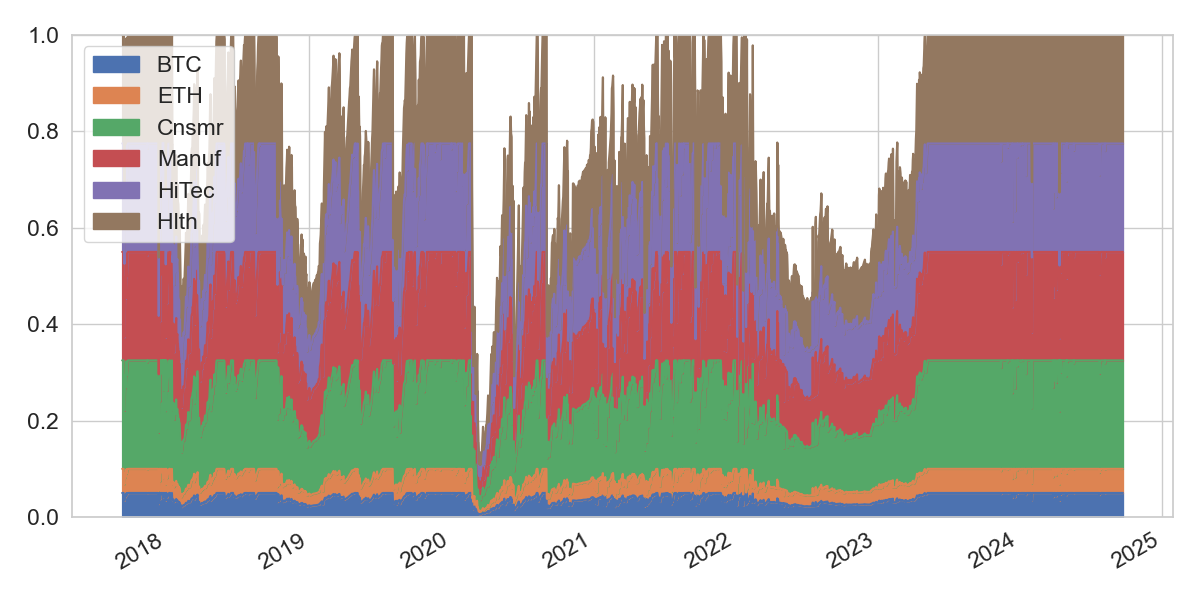}   
\caption{GARCH volatility estimator.}
\label{fig:weights_garch}
\end{subfigure}
\caption{Portfolio weights of the DD90/10 portfolios with EWMA and GARCH
volatility estimators.}
\label{fig:DD90_10_weights}
\end{figure}

\begin{figure}
\centering
\includegraphics[width=0.7\textwidth]{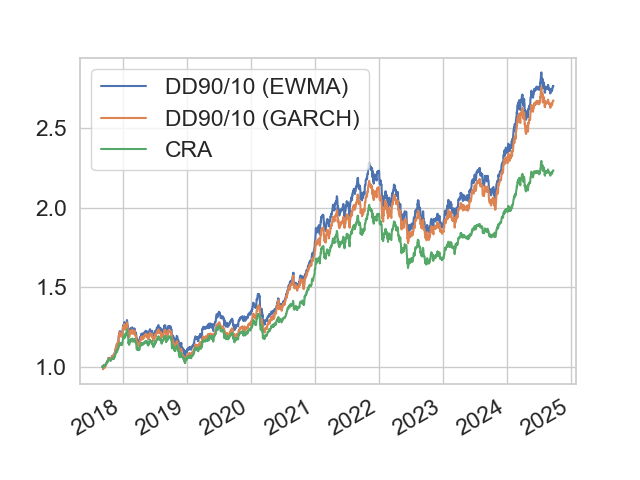}
\caption{Portfolio values of the DD90/10 and CRA portfolios.}
\label{fig:DD90-10_vs_CRA}
\end{figure}

\paragraph{Weights.} The weights of the DD90/10 portfolios are shown in
figure~\ref{fig:DD90_10_weights}.
The weights are quite similar, with the GARCH estimator being noticeably more
reactive during some periods. \paragraph{Performance.} The performance of the
DD90/10, compared to the (combined) CRA portfolio, is shown in
table~\ref{tab:DD90_10_stats}.
\begin{table}
\centering
\begin{tabular}{lrrr}
\toprule
Metric         & DD90/10 (EWMA)  & DD90/10 (GARCH) & CRA \\ 
\midrule
Return (\%)    & 10.4  & 10.1 & 8.2 \\
Volatility (\%)     & 9.8 & 9.7 & 8.2   \\
Sharpe        & 1.06  & 1.04 & 1.00 \\
Drawdown (\%)   & 19.9  & 19.7  & 19.6 \\ 
\bottomrule
\end{tabular}
\caption{Performance metrics DD90/10 and CRA.}
\label{tab:DD90_10_stats}
\end{table}
Figure~\ref{fig:DD90-10_vs_CRA} shows the value of the three portfolios over
time.

\section{Conclusions}
We have illustrated that, despite the documented extreme behaviors of crypto
assets, simple traditional portfolio contrstruction techniques can be used to
include them in a diversified portfolio.  We show this using two standard portfolio
construction methods, one based on risk parity, and the other a fixed set of relative 
weights, with each one dynamically diluted with cash to achieve a target ex-ante
risk.
The addition of even a modest crypto weight of 10\% 
increases the return and Sharpe ratio
of the portfolio significantly, without significantly increasing volatility or
drawdown. 

\paragraph{Acknowledgments.} This work was supported by the IOG Research Hub.

\clearpage
\bibliography{refs}
\end{document}